\documentclass[aps,prl,onecolumn,superscriptaddress,10pt,floatfix,amsmath,amssymb]{revtex4-2}

\usepackage{graphicx} 
\usepackage{dcolumn}
\usepackage{xcolor}
\usepackage{bm}
\usepackage{physics}
\usepackage{lipsum}  
\usepackage{hyperref}
\usepackage[normalem]{ulem}
\graphicspath{{images/}}

\begin{document}

\title{Assessing a binary quantum channel exploiting a Silicon photomultiplier based hybrid receiver}

\date{\today}

\author{Alberto Sanvito}
   \affiliation{University of Insubria, Via Valleggio 11, I-22100 Como (Italy)}
   
   \author{Silvia Cassina}  \email{s.cassina@uninsubria.it}
   \affiliation{University of Insubria, Via Valleggio 11, I-22100 Como (Italy)}
  
  \author{Marco Lamperti}
   \affiliation{University of Insubria, Via Valleggio 11, I-22100 Como (Italy)}
   \affiliation{Institute for Photonics and Nanotechnologies, IFN-CNR, Via Valleggio 11, I-22100 Como (Italy)} 
  
     \author{Michele N. Notarnicola}
   \affiliation{University of Milan and INFN Section of Milan, Via Celoria 16, I-20133 Milan (Italy)} 
   
   \author{Stefano Olivares}
   \affiliation{University of Milan and INFN Section of Milan, Via Celoria 16, I-20133 Milan (Italy)} 
   
   \author{Alessia Allevi}
   \affiliation{University of Insubria, Via Valleggio 11, I-22100 Como (Italy)}
   \affiliation{Institute for Photonics and Nanotechnologies, IFN-CNR, Via Valleggio 11, I-22100 Como (Italy)}

\begin{abstract}
In quantum communication protocols, the use of photon-number-resolving detectors could open new perspectives by broadening the way to encode and decode  information, and merging the properties of discrete and continuous variables. In this work, we consider a quantum channel exploiting a Silicon-photomultiplier-based receiver and evaluate its performance for quantum communication protocols under three possible configurations, defined by different post-processing of the detection outcomes. We investigate two scenarios: information transmission over the channel, quantified by the mutual information, and continuous-variable quantum key distribution. The preliminary results encourage further use of this detection scheme in extended networks.  
\end{abstract}

\maketitle

\section{Introduction}\label{sec:intro}
In the context of optical quantum communication, the encoding of information can be carried out either in discrete or continuous variables \cite{cariolaro}. While in the former case, the typical detection system is based on the use of single-photon detectors \cite{flamini}, in the latter one the signal detection is performed by means of homodyne detection schemes \cite{grosshans,diamanti, olivares}. In this second kind of detectors, the state to be characterized interferes at a balanced beam splitter with a high-intensity coherent state, the local oscillator (LO), and then the two outputs of the interferometer are detected by two photodiodes, whose photocurrent difference is measured as a function of the LO phase \cite{lvovsky}. Years ago, some of us have proposed to merge the two above-mentioned detection strategies in a hybrid scheme, in which photodiodes are substituted by photon-number-resolving (PNR) detectors \cite{donati}, and the LO has an energy comparable to that of the signal under investigation \cite{IJQI17}. The scheme has been already exploited to perform quantum state tomography \cite{NJP19} as well as coherent-state discrimination with binary phase-shift keying (BPSK) \cite{OE17}. Furthermore, theoretical works proved the adoption of PNR receivers for quantum communication protocols with coherent-state encoding to be beneficial, both for information
transmission \cite{martinez,ribeiro, lukanowski, jasc} and continuous-variable quantum key distribution (CVQKD) \cite{cattaneo,notarnicola23}. A first experimental verification has been recently carried out by DiMario and Becerra, who adopted a displacement PNR receiver, inspired on the Kennedy receiver for BPSK discrimination \cite{kennedy}, to assess the information transmission over realistic quantum channels, optimizing also the encoding strategy \cite{dimario, dimario1}. Anyway, the performance of a hybrid receiver, probing both the particle- and wave-like properties of radiation, has been so far limited to the quantum-state discrimination framework.\\
In this work, we extend the applicability of the hybrid scheme introduced in \cite{IJQI17} to the more general context of quantum communication with coherent states affected by loss. To this aim, we consider Silicon photomultipliers (SiPMs) \cite{OL19} as PNR detectors instead of the previously-used hybrid photodetectors \cite{JMO09}, since they are more compact and with higher dynamic range \cite{cassina21}. In particular, we evaluate the performance of the quantum channel exploiting the SiPM-based receiver under two different scenarios. At first, we address the transmission of classical information over the channel, quantified by the mutual information (MI), whilst, thereafter we consider CVQKD, proving conditional security under the wiretap channel assumption \cite{Pan}. In this latter framework, we compute the amount of secure information shared by Alice and Bob
both in direct (DR) and reverse reconciliation (RR) for individual and collective attacks. 
In both the use cases, we calculate the MI as a function of the parameters characterizing the system, and investigate the role of both the LO intensity \cite{PLAhomo}  and the losses affecting the encoded coherent states. 
In this way, we can characterize the communication channel and test its robustness to loss and detector inefficiencies.
Thanks to the features of our detection system, we consider three possible scenarios according to the output reading. In the first case we consider the detection system as a weak-field (WF) receiver, in which we have direct access to the output of the two employed PNR detectors. In the second case, we treat it as a homodyne-like (HL) receiver, where we evaluate the photon-number difference between the two outputs and use it to calculate the MI similarly to the procedure of a standard homodyne detection. Finally, in the third case, we employ a binary discrimination strategy (BDS), which is based on the sign of the photon-number difference in the case of a binary alphabet. 
In both applications, we consider BPSK encoding, providing the basic case for further implementation of more elaborated protocols with larger alphabets; we compare the experimental results to the corresponding figures of merit and demonstrate the feasibility of the proposed receiver for quantum communication protocols. 
Our results prove that, in the presence of PSK modulation, the WF and HL strategies are equivalent, while the BDS method is less performing, according to the data processing inequality. 
The obtained outcomes encourage further exploitation of the SiPM-based detection scheme in extended networks \cite{notarnicola}. 

\section{Theoretical description}\label{theory}

In this work, we address a BPSK protocol, being one of the basic strategies to transmit information between two distant parties: a sender, usually called Alice, and a receiver, usually called Bob \cite{cariolaro,helstrom,bergou}. BPSK can be easily implemented by encoding two classical symbols $k=0,1$ onto the coherent states:
\begin{align}
    |\alpha_k\rangle = |e^{i\pi (k+1)} \alpha\rangle \, , \qquad (k=0,1) \, ,
\end{align}
with $\alpha>0$, such that $|\alpha_0\rangle=|-\alpha \rangle$ and $|\alpha_1\rangle=|+\alpha \rangle$, generated with equal a priori probabilities $q_k=1/2$. The encoded pulses have the same energy, i.e. mean number of photons, equal to $\alpha^2$, but are phase-shifted by $\pi$ \cite{izumi,muller}. We note that the choice of coherent-state encoding provides robustness against losses, since with respect to nonclassical states of radiation, coherent states remain coherent under a lossy dynamics \cite{Olivares2021}.
The encoded pulses are then injected into the quantum channel, here considered as a pure-loss channel associated with transmissivity $T<1$, until to reach Bob, who performs a suitable measurement on the received signal $|\sqrt{T} \alpha_k \rangle$.\\
The adopted detection scheme is based on an interferometric setup, where the received quantum state $|\sqrt{T} \alpha_k \rangle$ interferes at a balanced beam splitter (BS) with a low-intensity LO prepared in the coherent state $|z\rangle$ with amplitude $z>0$. Thereafter, we perform PNR detection at each BS output, and reconstruct the local photon-number distributions on both arms. In particular, if state $|\alpha_k\rangle$ is sent, the conditional photon-number distributions of the transmitted and reflected beams are given by Poisson distribution, namely:
\begin{align}
{\cal P}_n (\mu|\alpha_k) = e^{-\mu} \frac{\mu^n}{n!} \, ,
\end{align}
with rates $\mu_k^{(t(r))}$, respectively, in which
\begin{align}\label{eq:mutr}
    \mu_k^{(t)} = \frac12 \left( T \alpha^2 + z^2 + 2 \xi \, \sqrt{T} \alpha_k  z \right) \quad \mbox{and} \quad  \mu_k^{(r)}= \frac12 \left( T \alpha^2 + z^2 - 2 \xi \, \sqrt{T} \alpha_k  z \right)  \, ,
\end{align}
$\xi \le 1$ representing the interference visibility, quantifying the overlap between the signal and the LO beams interacting with each other \cite{notarnicola, dimario}.


Given the previous setup, we identify three different strategies, that, ultimately, correspond to three different types of receivers.
In the former, referred to as weak-field (WF) receiver, we have access to the single detector outputs $n$ and $m$ on the transmitted and reflected arm of the BS, respectively. Therefore, the outcome of the measurement is provided by the tuple of integers $(n,m)$. In turn, the conditional probability given state $|\alpha_k\rangle$ is the bi-variate Poisson distribution:
\begin{align}\label{pwf}
p_{\rm WF}(n,m|\alpha_k) = {\cal P}_n \left(\mu_k^{(t)} \big|\alpha_k\right) {\cal P}_m \left(\mu_k^{(r)} \big|\alpha_k\right)\, .
\end{align}
In the second strategy, referred to as homodyne-like (HL) receiver, we use the single PNR outputs to only evaluate, in post-processing, the photon-number difference, $\Delta=n-m$, $\Delta \in Z$. In turn, the probability distribution $p_{\rm HL}(\Delta|\alpha_k)$ follows a Skellam distribution \cite{OE17}:
\begin{align}\label{skellam}
    p_{\rm HL}(\Delta|\alpha_k) =& \sum_{m=0}^{\infty} {\cal P}_{m+\Delta} \left(\mu_k^{(t)} \big|\alpha_k\right) {\cal P}_m \left(\mu_k^{(r)} \big|\alpha_k\right) \nonumber \\
=& e^{-\mu_k^{(t)} - \mu_k^{(r)}} \left(\mu_k^{(t)}\right)^{\Delta} \sum_{m=0}^{\infty} \frac{\left(\mu_k^{(t)} \mu_k^{(r)}\right)^m}{m!(\Delta+m)!}\nonumber\\
    =& e^{-\mu_k^{(t)} - \mu_k^{(r)}} \left[ \frac{\mu_k^{(t)}}{\mu_k^{(r)}} \right]^{\Delta/2} I_{\Delta}\left(2 \sqrt{\mu_k^{(t)} \mu_k^{(r)}} \right) \, ,
\end{align}
where $I_{\Delta}(x)$ is the modified Bessel function of the first kind. 
Moreover, as demonstrated in \cite{OE17, jasc}, the Skellam distribution converges to a homodyne (Gaussian) distribution in the limit of macroscopic LO, $z^2 \to \infty$. 
Finally, the last strategy requires further post-processing on the HL data, drawing inspiration on the methods of quantum-state discrimination theory, and, thus, referred to in the following as binary discrimination strategy (BDS) \cite{cariolaro, bergou, becerra,becerra1}. In more detail, in the BDS we design a binary positive-operator-valued measurement (POVM), retrieving the two integer outcomes $j=0,1$, that directly infers which was the state $|\alpha_k\rangle$ generated by the sender. This binary POVM is constructed by evaluating the sign of the measured photon-number difference $\Delta=n-m$. In particular, given the phase space representation associated with BPSK encoding, we set outcome $j=0$ if a negative difference $\Delta<0$ is obtained, and $j=1$ if the measured difference is positive, $i.e.$ $\Delta>0$, while, in the case $\Delta=0$, we perform a random decision between $j=0$ and $j=1$, with equal probability.
The corresponding conditional probability $p_{\rm BDS}(j|\alpha_k)$ then reads:
\begin{align}
p_{\rm BDS}(0|\alpha_k) &= \sum_{\Delta<0} p_{\rm HL}(\Delta|\alpha_k) + \frac{p_{\rm HL}(\Delta=0|\alpha_k)}{2}   \nonumber \, \\[1ex] 
p_{\rm BDS}(1|\alpha_k) &= 1- p_{\rm BDS}(0|\alpha_k) \, .
\end{align}
We also note that, due to the non-orthogonality of the encoded coherent states, the distribution $p_{\rm BDS}(j|\alpha_k)$ is nontrivial, since there is always a nonzero probability to infer the wrong symbol, i.e. $p_{\rm BDS}(j\ne k|\alpha_k) \ne 0$, which, in the context of BPSK discrimination, leads to a nonzero decision error probability \cite{cariolaro, bergou,becerra,becerra1}.

In the following, we address the performance of these three strategies in two different quantum communication contexts. At first, we address a genuine communication protocol and quantify the information transmission between Alice and Bob over the proposed channel. Thereafter, we consider possible applications to CVQKD.

\subsection{Information transmission over the channel}

To begin with, we compute the information rate transmitted over the channel, quantified by the MI, which yields the amount of information encoded at the source that Bob is effectively able to extract from his probed statistics.
In general, given two stochastic variables $A$ and $B$, associated with probabilities $p_A(x_A)$, $x_A \in A$, and $p_B(x_B)$, $x_B\in B$, and a communication channel $X\to Y$ described by the conditional probability distribution $p_{B|A}(x_B|x_A)$, the MI is defined as \cite{cover}:
\begin{align}\label{MIdef}
    I(A;B) = H(B) - H(B|A) \, ,
\end{align}
expressed in bits per channel use,
where $H(B) = {\sf H}[p_B(x_B)]$ is the overall entropy of $B$, with $p_B(x_B)=\sum_{x_A \in A} p_A(x_A)\\ p_{B|A}(x_B|x_A)$, and $H(B|A) = \sum_{x_A \in A} p_A(x_A) {\sf H}[p_{B|A}(x_B|x_A)]$ is the average conditional Shannon entropy of $B$ if $A$ is given, in which 
\begin{align}
{\sf H}[p(x)]= - \sum_x p(x) \log_2 p(x)
\end{align}
represents the Shannon entropy of a probability distribution $p(x)$ \cite{cover}.

Keeping this in mind, we now compare the MI for the three methods under the assumption that Alice encodes binary information on symbols $k=0,1$, with equal a priori probability $q_k=1/2$, while Bob performs different measurement strategies, associated with different conditional probabilities, and different amounts of MI.
In more detail, for the WF receiver we have:
\begin{align}\label{MIWF}
I_{\rm WF}(A;B) &= {\sf H} \left[ \sum_{k=0,1} q_k p_{\rm WF}(n,m|\alpha_k) \right] - \sum_{k=0,1} q_k \,{\sf H} \left[ p_{\rm WF}(n,m|\alpha_k) \right] \nonumber \\[1ex]
&=- \sum_{n,m=0}^{\infty} \left( \sum_k q_k p_{\rm WF}(n,m|\alpha_k) \right) \log_2 \left( \sum_k q_k p_{\rm WF}(n,m|\alpha_k)\right) \nonumber \\[1ex]
& \hspace{1.5cm} +\sum_{k} q_k \left[\sum_{n,m=0}^{\infty} p_{\rm WF}(n,m|\alpha_k) \log_2  p_{\rm WF}(n,m|\alpha_k) \right]\, ,
\end{align}
while for the HL:
\begin{align}\label{MIHL}
I_{\rm HL}(A;B) &= {\sf H} \left[ \sum_{k=0,1} q_k p_{\rm HL}(\Delta|\alpha_k) \right] - \sum_{k=0,1} q_k \,{\sf H} \left[  p_{\rm HL}(\Delta|\alpha_k) \right] \nonumber \\[1ex]
&=- \sum_{\Delta=-\infty}^{\infty} \left( \sum_k q_k  p_{\rm HL}(\Delta|\alpha_k) \right) \log_2 \left( \sum_k q_k  p_{\rm HL}(\Delta|\alpha_k)\right) \nonumber \\[1ex]
& \hspace{1.5cm} +\sum_{k} q_k \left[\sum_{\Delta=-\infty}^{\infty}  p_{\rm HL}(\Delta|\alpha_k) \log_2   p_{\rm HL}(\Delta|\alpha_k) \right]\, ,
\end{align}
and, eventually, for the BDS:
\begin{align}\label{BDS}
I_{\rm BDS}(A;B) &= {\sf H} \left[ \sum_{k=0,1} q_k p_{\rm BDS}(j|\alpha_k) \right] - \sum_{k=0,1} q_k \,{\sf H} \left[  p_{\rm BDS}(j|\alpha_k) \right] \nonumber \\[1ex]
&=- \sum_{j=0,1} \left( \sum_k q_k  p_{\rm BDS}(j|\alpha_k) \right) \log_2 \left( \sum_k q_k  p_{\rm BDS}(j|\alpha_k)\right) \nonumber \\[1ex]
& \hspace{1.5cm} +\sum_{k} q_k \left[\sum_{j=0,1}  p_{\rm BDS}(j|\alpha_k) \log_2  p_{\rm BDS}(j|\alpha_k) \right]\, .
\end{align}
Given these expressions, it is 
worth to investigate which strategy 
yields the highest value of MI. Therefore, we remind a fundamental result of information theory, namely the data processing inequality, stating that any post-processing of data implies a loss of bits, and thus a reduction of MI \cite{cover, jasc}. Thus, in our scenario we expect the following hierarchy to hold: $I_{\rm WF}(A;B) \ge I_{\rm HL}(A;B) \ge I_{\rm BDS}(A;B)$. 
Nevertheless, in the presence of PSK modulation, the WF and HL are equivalent in terms of MI. 
In fact, implementing WF detection, retrieving the outcome $(n,m)$, is equivalent to perform joint measurement of both the sum and difference photocurrents $\sigma=n+m$, and $\Delta=n-m$, respectively, with $\sigma\in \mathbb{N}$ and $\Delta \in \mathbb{Z}$. In turn, the joint probability $P(\sigma,\Delta|\alpha_k)$ contains the same amount of information on the encoded signal as $p_{\rm WF}(n,m|\alpha_k)$. Moreover, the distribution can be re-expressed as $P(\sigma,\Delta|\alpha_k)= p_{\rm HL}(\Delta|\alpha_k) f(\sigma, \Delta)$, for a suitable function $f(\sigma,  \Delta)$, being independent of $\alpha_k$ and satisfying $\sum_\sigma f(\sigma,  \Delta)=1$. Accordingly, information on the signal phase is only carried by the difference photocurrent, and, after straightforward calculation, we get  $I_{\rm WF}(A;B)=I_{\rm HL}(A;B)$.\\
However, in practical implementations, one should also design suitable reconciliation codes to practically extract a finite number of bits from the output statistics of the receivers, which would introduce further loss of bits, thus making the analysis of suboptimal methods like BDS still worth of investigation.

\section{Experimental implementation}
\subsection{Setup and preliminary characterization}
\begin{figure}[htbp]
\centering\includegraphics[width=8cm]{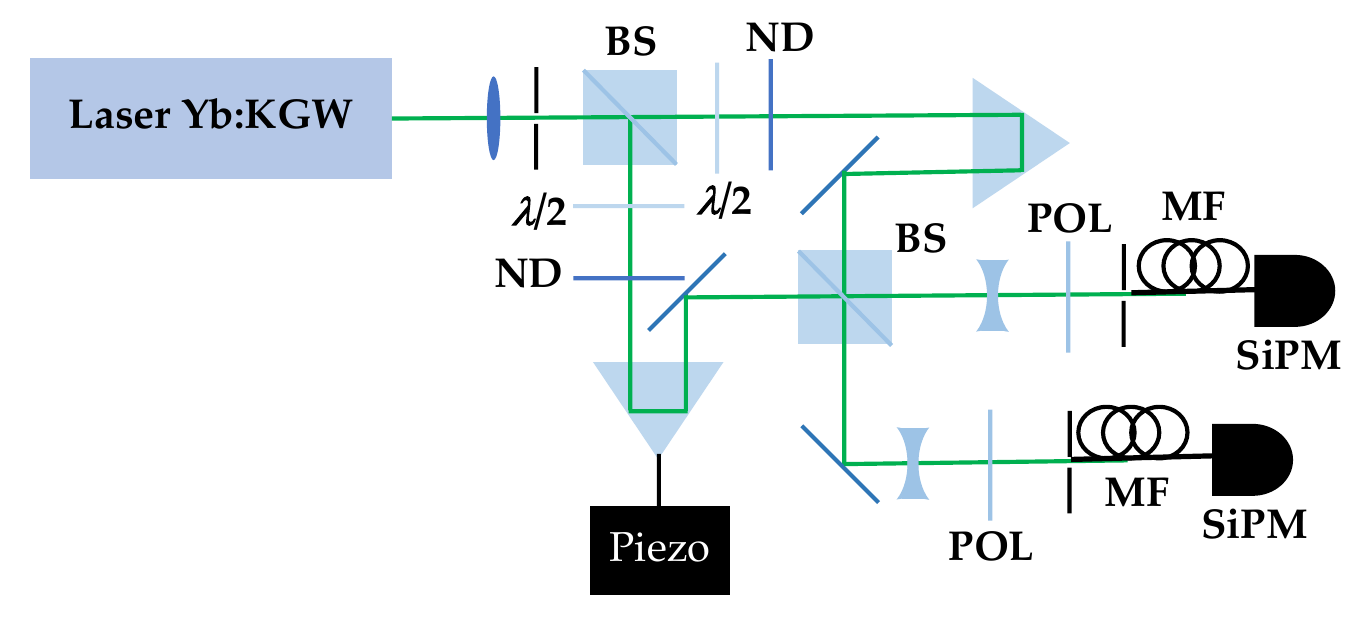}
\caption{Sketch of the experimental setup to realize the SiPM-based receiver: BS, beam splitter; $\lambda/2$, half-wave plate; ND, variable neutral-density filter; Piezo, piezoelectric mounted on a translation stage; POL, polarizer; MF, multi-mode fiber; SiPM, Silicon photomultiplier.} \label{setup}
\end{figure}

In this paragraph, we give a more detailed description of the detector based on SiPMs.
In particular, we are interested in demonstrating that such a scheme can be embedded in
quantum communication protocols, both for information transmission and CVQKD.\\
The experimental setup is shown in Fig.~\ref{setup}. The second-harmonic pulses (at 515 nm, 190 fs pulse duration)
of a Yb:KGW laser operated at 5 kHz are sent to a Mach-Zehnder interferometer, whose input BS divides the light beam into two parts: in one arm we have the signal, while in the other arm the LO. The length of one arm of the interferometer is controlled in steps by means of a piezoelectric movement (Pz) in order to modify the LO phase through the whole $2 \pi$ range. Indeed, the selection of the states $|\alpha \rangle$ and $|-\alpha \rangle$ is obtained in post-processing by the comparative analysis of the mean value of the light as a function of the LO phase and of the photon-number difference distribution, as already explained in Refs.~\cite{IJQI17,OE17}. This operation is performed for different attenuations of the signal and LO obtained by means of two variable neutral density (ND) filters. 
To optimize the overlap between the signal and the LO beams in the second BS of the interferometer, we worked on the control of the polarization of the beams as well as on their spatio-temporal superposition. In particular, we reduced the beam size to make its attenuation uniform on the variable density filter. Concerning the temporal superposition, we used a piezoelectric movement of the translation stage in one arm of the interferometer to optimize it (see Fig.~\ref{setup}).\\
At the two outputs of the second BS two collection systems are placed. They are formed by a pin-hole with a fixed aperture and a multi-mode fiber with a 1 mm core that delivers the light to two SiPMs. The model that we use is the MPPC S13360-1350CS produced by Hamamatsu Photonics \cite{hama}, which consists of 667 pixels in a 1.3 $\times$ 1.3 mm$^2$ photosensitive area, with a pixel pitch equal to 50 $\mu$m. This kind of detector is endowed with a good photon-number-resolving capability and is operated at room temperature. The output of each detector is amplified by a fast inverting amplifier with a gain of 24 dB embedded in the computer-based Caen SP5600 Power Supply and Amplification Unit, synchronously integrated by a boxcar-gated integrator, and digitized. We choose an integration gate of 15 ns centered around the peak of the amplified signal \cite{OL19}. For each condition of signal and LO, 10$^5$ pulses are acquired. 
\begin{figure}[htbp]
\centering\includegraphics[width=8cm]{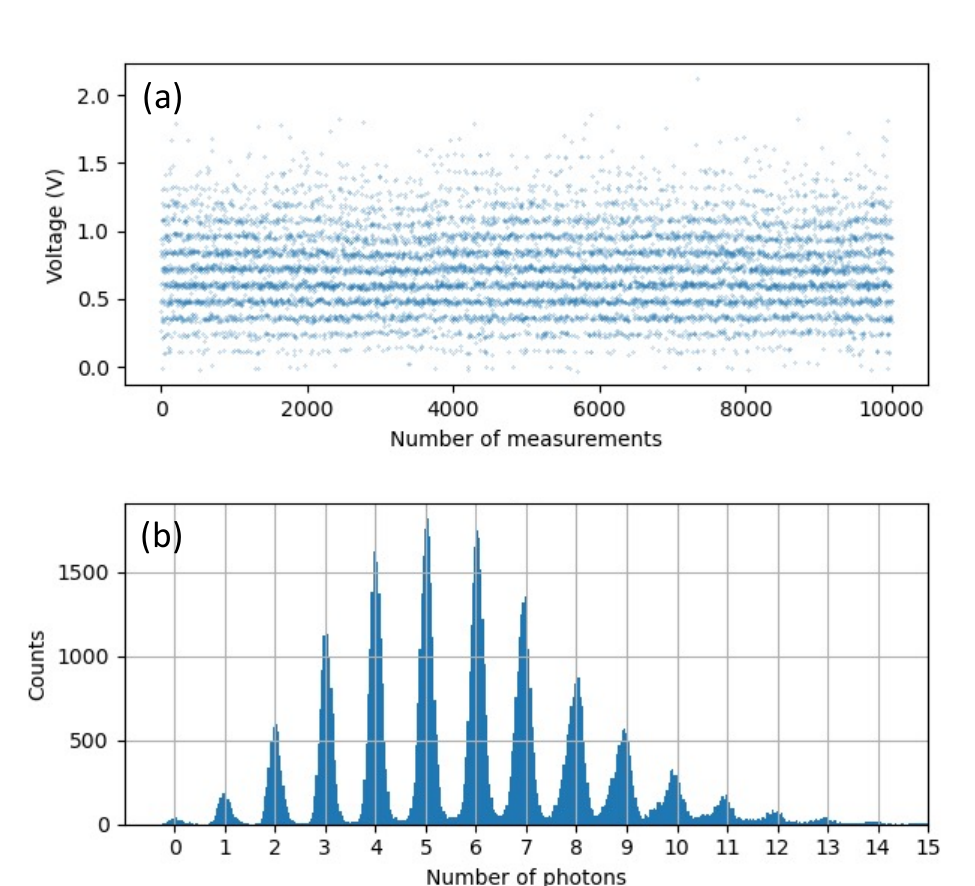}
\caption{(a) Typical trace of $10^4$ shots acquired by a SiPM embedded in the hybrid scheme; (b) example of pulse-height spectrum corresponding to a mean value $\langle m \rangle = 5.77$.} \label{output}
\end{figure}
A typical trace, for a given value of signal and LO, is shown in Fig.~\ref{output}(a), while a typical reconstructed pulse-height spectrum (PHS) is shown in Fig.~\ref{output}(b) \cite{cassina21}. 
\subsection{Experimental results on information transmission}
\begin{figure}[htbp]
\centering\includegraphics[width=8cm]{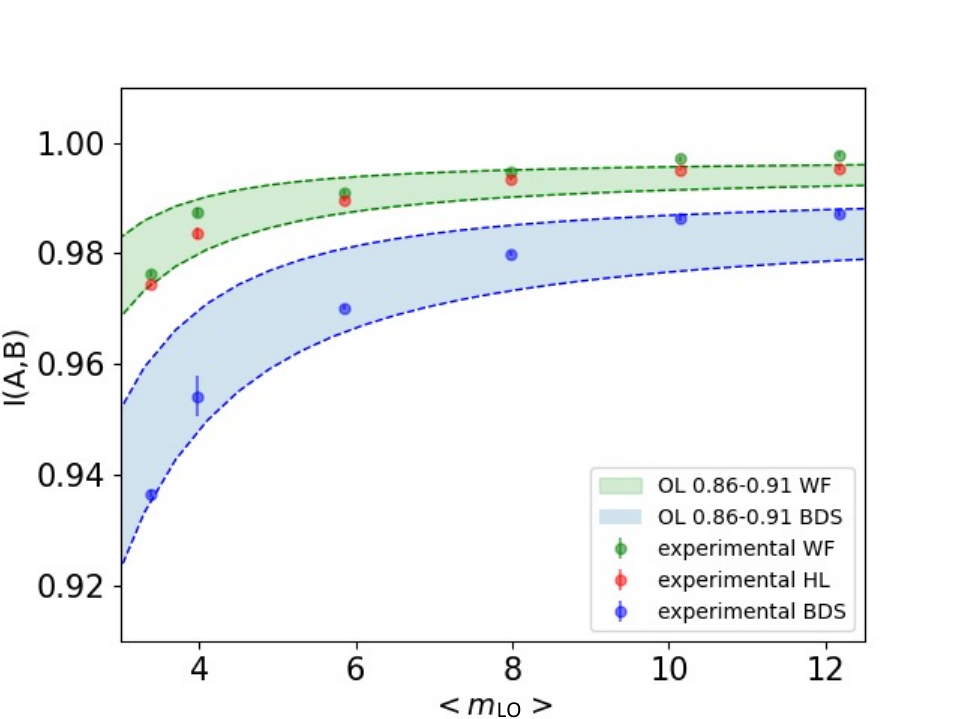}
\caption{MI as a function of the mean value of LO, for a fixed mean value of the signals, $\langle m_{\rm sig}\rangle = 3.07$: green dots refer to the WF receiver, red dots to the HL receiver, and blue dots to the BDS receiver. Dots + error bars: experimental data;  dashed lines: results from numerical expectations, in which we assumed $\xi = 0.86$ (lower curve) and $\xi = 0.91$ (higher curve). The highlighted bands were calculated according to Eqs.~(\ref{MIWF})-(\ref{BDS}) with an overlap ranging from $\xi=0.86$ to $\xi=0.91$.} \label{MIvsLO}
\end{figure}

The characterization of the SiPM-based receiver used for state discrimination can be obtained as a function of some important parameters, namely the energy of the LO and the losses affecting the signal. 
In particular, in Fig.~\ref{MIvsLO} 
we consider the behavior of $I(A;B)$ as a function of the mean number of LO photons for a fixed mean number of signal photons ($\langle m_{\rm sig}\rangle = 3.07$). The data obtained from the WF receiver are shown as green dots, those from the HL receiver as red dots, and those from the BDS receiver as blue dots. We immediately note that the first two methods give the same results since the corresponding data are perfectly superimposed, while those corresponding to BDS are lower. The experimental results are superimposed to the numerical expectations (see the highlighted bands in the same figure), calculated according to Eqs.~(\ref{MIWF})-(\ref{BDS}) with an overlap ranging from $\xi=0.86$ to $\xi=0.91$. The colored dashed curves represent the theoretical expectation limits for $\xi=0.86$ (lower curve) and $\xi=0.91$ (upper curve). The variation of $\xi$ is caused by the fact that changing the LO intensity by means of the ND filter could slightly change the overlap between the interfering beams at the BS. Despite these imperfections, from the analysis of Fig.~\ref{MIvsLO} we can observe that MI reaches values larger than 0.98 quite rapidly. Indeed, the asymptotic value that can be achieved for $\langle m_{\rm LO} \rangle \rightarrow \infty$ is equal to 1 in the case of BPSK protocols. Experimentally, the maximum value of MI is obtained for the highest mean value we measured, namely $\langle m_{\rm LO} \rangle = 12.17$.\\
As a second step of investigation, we consider the effect of losses affecting the signal. To this aim, we set the mean value of LO at $\langle m_{\rm LO} \rangle = 12.15$, and vary the energy of the signal by means of the variable ND filter shown in Fig.~\ref{setup} from $\langle m_{\rm sig}\rangle = 3.20$ to $\langle m_{\rm sig}\rangle = 0.14$, thus considering a maximum $13.44$ dB loss. 
\begin{figure}[htbp]
\centering\includegraphics[width=8cm]{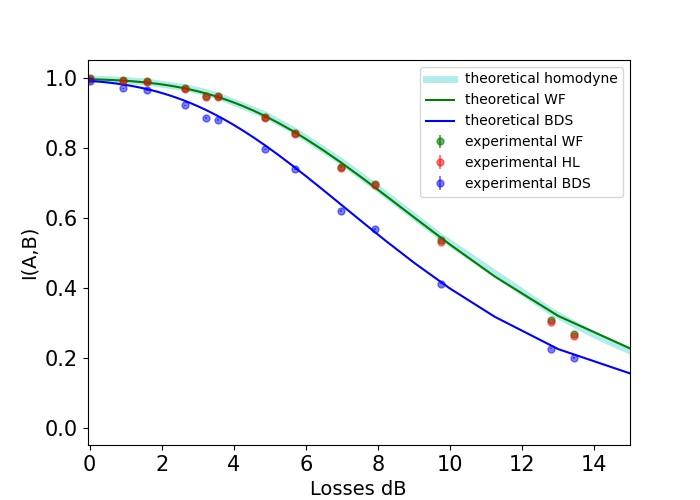}
\caption{MI as a function of the losses affecting the signal arm, for a fixed mean value of the LO, $\langle m_{\rm LO} \rangle = 12.15 \sim 4 \langle m_{\rm sig}\rangle_{\rm MAX}$. Green dots refer to the WF receiver, red dots to the HL receiver, and blue dots to the BDS receiver. Dots + error bars: experimental data; colored lines: results from numerical expectations (HL and WF correspond to the same curve); cyan line: theoretical prediction of the standard homodyne receiver. In the numerical and homodyne predictions we assumed $\xi = 0.94$.} \label{MIvsLOSS}
\end{figure}
The experimental results corresponding to MI are shown in Fig.~\ref{MIvsLOSS} as functions of the losses affecting the signal. The experimental data are shown as colored dots, respectively, while the numerical expectations are presented as solid curves with the same color choice. As in the case of Fig.~\ref{MIvsLO}, green dots refer to the WF receiver, red dots to the HL receiver, and blue dots to the BDS receiver. Again, the agreement between the experimental data and numerical expectations is achieved assuming a non-perfect overlap, that in this kind of investigation is $\xi=0.94$.
While we observe that there is a perfect overlap between the data corresponding to WF and HL methods (and the same holds for the theoretical expectations), lower values of MI are obtained through the BDS strategy, as already discussed in the previous Section.  
Moreover, in the same figure we plot, as cyan solid curve, the expected results obtained from a standard homodyne receiver, having the same overlap as our experimental receiver. We can notice that the performance of the two detection systems is quite similar, despite the fact that our detector is based on discrete quantities (numbers of photons), while standard homodyne detection relies on continuous variables. This means that the proposed receiver can be considered as a valid alternative to the more common technique, having the advantage of a low-energy LO.
\subsection{Towards applications in continuous-variable quantum key distribution}

The interest in investigating a SiPM-based detection scheme can be further extended to the field of CVQKD, with nontrivial consequences. 
In fact, it has been recently demonstrated that the optimal strategies for state discrimination, information transmission and CVQKD differ from one another, even if based on the same kind of detection system \cite{notarnicola23}.\\ 
Let us consider the communication channel between Alice and Bob to be eavesdropped by a third party, Eve. Alice and Bob's goal is to share a common random secure key. In the simplest case, Alice generates randomly one of the two symbols $k=0,1$, with corresponding probability $q_k=1/2$, and Bob retrieves a set of data correlated to Alice's ones. 
The figure of merit is now the KGR, defined as the difference between the mutual information between Alice and Bob and the one acquired by Eve.  
We investigate security by considering both individual attacks (IA) and collective attacks (CA). In the former case, the eavesdropper measures individually each sent pulse, in the latter a final joint collective measurement is performed over many transmitted signals according to either DR or RR, respectively \cite{pirandola1}.
In general, the security can be addressed under different paradigms, according to the level of trust of the quantum channel \cite{pirandola}. Here, we assume a restrictive eavesdropping scenario under a wiretap channel. In other words, we assume the transmission channel to be completely characterized in terms of a pure-loss channel, so that Eve has only access to the fraction of the signals lost during transmission, without either performing any arbitrary channel manipulation or injecting noise in the system \cite{Pan, notarnicola23}. Given this consideration, when Alice sends the state $|\alpha_k\rangle$, Bob and Eve receive the transmitted and reflected fractions  $|\sqrt{T}\alpha_k\rangle$ and $|\sqrt{1-T}\alpha_k\rangle$, respectively.\\
We start with the IA scenario for the receiver that guarantees higher values of MI, that is the WF receiver. In this case, we assume Eve to hold the same receiver as Bob in its best performing solution. Thus, Eve has at her disposal a WF receiver with unit visibility $\xi_E=1$. For the sake of simplicity, in this scenario we also consider Bob to hold a WF receiver, albeit with reduced visibility $\xi_B <1$. The corresponding KGR reads
\begin{align}
   \Delta I_{\rm WF}^{\rm (IA)} = I_{\rm WF}(A;B)- I_{\rm WF}(A;E) \qquad \mbox{for DR} \, ,
\end{align}
with $I_{\rm WF}(A;E)$ computed according to Eq.~(\ref{MIWF}), with the substitutions $T \to 1-T$ and $\xi_E=1$, and:
\begin{align}
   \Delta I_{\rm WF}^{\rm (IA)} = I_{\rm WF}(A;B)- I_{\rm WF}(B;E) \qquad \mbox{for RR} \, ,
\end{align}
in which the MI $I_{\rm WF}(B;E)$ is derived from the joint probability distribution of Alice, Bob and Eve:
\begin{align}
p_{ABE}(k; n_B,m_B; n_E, m_E) = q_k p_{\rm WF}^{(B)}(n_B,m_B|\alpha_k) p^{(E)}_{\rm WF}(n_E, m_E|\alpha_k) \, , 
\end{align}
in which $p_{\rm WF}^{(B(E))}$ is given in~(\ref{pwf}) with the parameters $T$ and $\xi_B$ for Bob's side, and $1-T$ and $\xi_E=1$ for Eve's.

On the other hand, in the presence of CA, the MI at Eve's side should be replaced by the Holevo information $\chi(A;E)$ and $\chi(B;E)$ for DR and RR, respectively, providing the maximum amount of information extractable from the ensemble of eavesdropped signals compatible with quantum mechanics laws.
However, since DR is inevitably associated with a bound of $3$ dB transmission (see the next Section), due to the symmetry of the channel, here we only consider the RR case, providing a more effective solution. Then, we have:
 \begin{align}
   \Delta I_{\rm p}^{\rm (CA)}= I_{\rm p} (A;B)- \chi_{\rm p} (B;E)\,,  \qquad ({\rm p= WF,HL, BDS})\, , 
\end{align}
where
\begin{align}
 \chi_{\rm p} (B;E) = S(E)- S_{\rm p} (E|B) 
\end{align}
is the Holevo information between Bob and Eve \cite{notarnicola23}. In the former expression, $S(E)=S[\rho_E]$, where $\rho_E= \sum_k q_k |\sqrt{1-T} \alpha_k\rangle \langle \sqrt{1-T} \alpha_k|$, and $S[\rho]=- {\rm Tr}[\rho \log_2 (\rho)]$ is the Von-Neumann entropy of the state $\rho$, whereas $S_{\rm p} (E|B) $ is Eve's entropy conditioned to Bob. In particular, for the WF case we have:
\begin{align}
S_{\rm WF} (E|B) = \sum_{n_B,m_B} p_{\rm WF}^{(B)}(n_B,m_B) \, S[\rho_{E|(n_B,m_B)}] \, ,
\end{align}
in which $p_{\rm WF}^{(B)}(n_B,m_B)= \sum_k q_k p_{\rm WF}^{(B)}(n_B,m_B|\alpha_k)$ is the overall distribution of the WF receiver, and $\rho_{E|(n_B,m_B)}$ is the corresponding conditional state received by Eve \cite{cattaneo, notarnicola23}. Also in this context, we show that WF and HL are again equivalent, as $\rho_{E|(n_B,m_B)}= \rho_{E|\Delta}$, with $\Delta=n_B-m_B$, thus $S_{\rm WF} (E|B)= S_{\rm HL} (E|B)$, while the conditional entropy in the presence of BDS is:
\begin{align}
S_{\rm BDS} (E|B) = \sum_{j=0,1} p_{\rm BDS}(j) \, S[\rho_{E|j}] \, ,
\end{align}
with $p_{\rm BDS}(j)= \sum_k q_k p_{\rm BDS}(j|\alpha_k)$.

\subsection{Experimental results}
\begin{figure}[htbp]
\hskip -0.8cm
\centering\includegraphics[width=14cm]{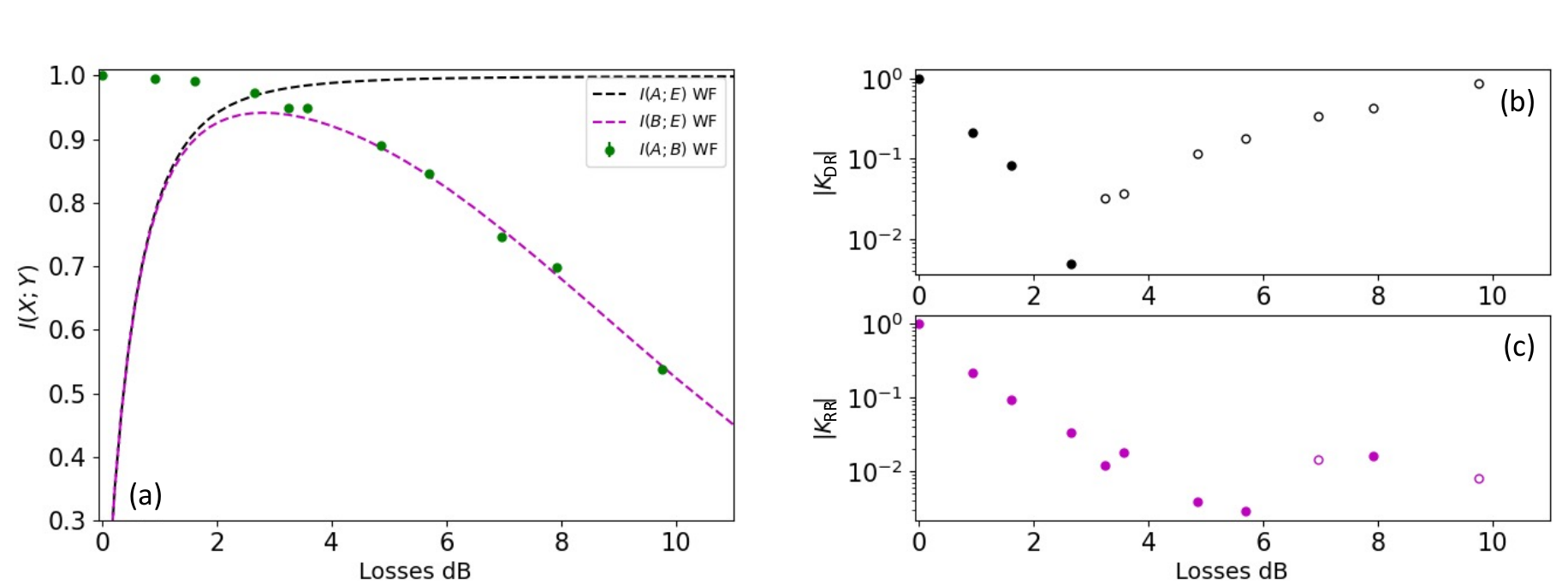}
\caption{IA scenario. (a) Mutual information $I_{\rm WF}(A; B)$ (experimental data, green dots), $I_{\rm WF}(A; E)$ (numerical expectation, black curve) and $I_{\rm WF}(B; E)$ (numerical expectation, magenta curve) as functions of the losses affecting the signal arm, for a fixed mean value of the LO, $\langle m_{\rm LO} \rangle = 12.15$, in a semi-logarithmic scale. In the numerical expectations we assumed $\xi_{\rm B} = 0.94$ and $\xi_{\rm E} = 1$. (b) and (c): $|K_{\rm DR}|$ and $|K_{\rm RR}|$ as functions of the losses affecting the signal arm, respectively. Full (open) symbols refer to positive (negative) values of $K_{\rm DR}$ and $K_{\rm RR}$.} \label{individual}
\end{figure}

\begin{figure}[htbp]
\hskip -0.8cm
\centering\includegraphics[width=14cm]{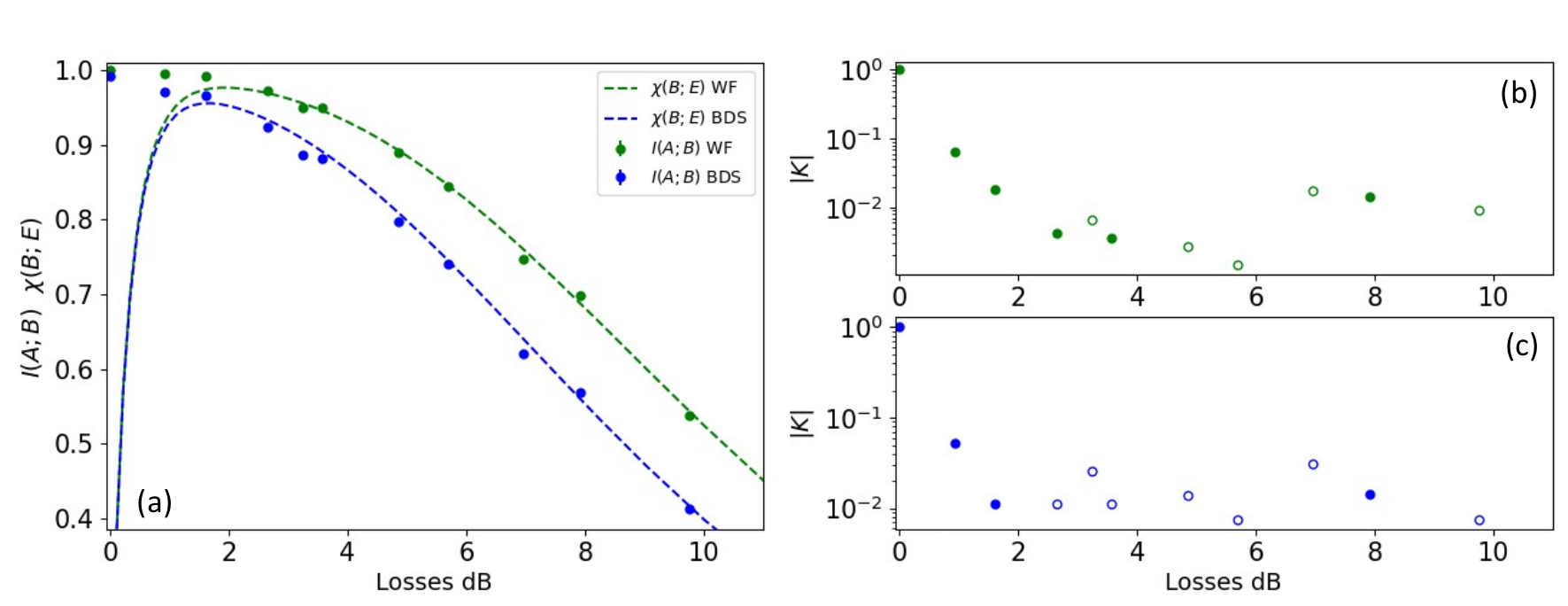}
\caption{CA scenario. (a) Mutual information $I_{\rm WF}(A; B)$ and $I_{\rm BDS}(A; B)$ (experimental data, green and blue dots, respectively), $\chi_{\rm WF}(A; E)$ and $\chi_{\rm BDS}(A; E)$ (numerical expectations, green and blue curve, respectively) as functions of the losses affecting the signal arm, for a fixed mean value of the LO, $\langle m_{\rm LO} \rangle = 12.15$, in a semi-logarithmic scale. In the numerical expectations we assumed $\xi_{\rm B} = 0.94$ and $\xi_{\rm E} = 1$. (b) and (c) $|K|$ as a function of the losses affecting the signal arm for WF and BDS receiver, respectively. Full (open) symbols refer to positive (negative) values of $K$.} \label{collective}
\end{figure}

In Fig.~\ref{individual}, 
we show the MI as a function of channel transmittance in the presence of individual attacks. For the sake of completeness, we consider both DR and RR strategies, and we assume that both Bob and Eve employ PNR detectors, and in particular WF method. However, to test the possible advantages of Eve over Bob, we consider a perfect overlap for the eavesdropper compared to the non-perfect overlap ($\xi = 0.94$) corresponding to Bob. Moreover, we assume that all the losses experienced by Bob correspond to the signal intercepted by Eve. In panel (a) we compare the experimental values of $I_{\rm WF}(A; B)$ with the numerical expectations of $I_{\rm WF}(A; E)$, which correspond to the case of DR, and those of $I_{\rm WF}(B; E)$, which correspond to the case of RR. In particular, from the figure we can easily appreciate for which values the MI between Alice and Bob surpasses that between Alice (or Bob) and Eve. As expected, the more effective strategy is to use RR instead of DR. Indeed, the communication rapidly drops in the case of DR, while it decreases more smoothly in the case of RR. In panels (b) and (c) of the same figure, we show the modulus of the quantities $K_{\rm DR}$ and $K_{\rm RR}$, respectively, defined as the normalized information between Alice and Bob in the presence of individual attacks, namely
\begin{eqnarray}\label{Kdrandrr}
K_{\rm DR} = \frac{ I_{\rm WF}(A; B)- I_{\rm WF}(A; E)}{I_{\rm WF}(A; B)} \nonumber\\
K_{\rm RR} = \frac{I_{\rm WF}(A; B)- I_{\rm WF}(B; E)}{I_{\rm WF}(A; B)}.
\end{eqnarray}
We point out that the full circles correspond to positive values of $K_{\rm DR}$ and $K_{\rm RR}$, while open symbols to negative ones.
Indeed, the calculation of the quantities in Eqs.~(\ref{Kdrandrr}) starting from real data is quite delicate when the absolute values become small, namely smaller than $10^{-3}$. In that case, the effect of noise due to the detection system, such as the cross talk of SiPM or the electronic noise of the boxcar-gated integrators \cite{OL19}, could play a detrimental role, which is instead not visible in the pure calculation of MI.\\
As a further investigation, we consider collective attacks. In this case, we have to consider the calculation of the Holevo bound, which is an upper limit for MI. Indeed, this definition is useful to define a lower bound for the secret key rate. By following Ref.~\cite{cattaneo}, we consider the asymptotic limit in which infinitely many signals are shared between Alice and Bob, thus neglecting any possible finite-size effect arising with finite datasets. Similarly to the case of individual attacks, also for collective ones we exploit the experimental data for the calculation of the MI between Alice and Bob, while we use the numerical expectation for the calculation of the Holevo bound. In particular, we only consider RR since it is more effective than DR. The results are shown in Fig.~\ref{collective} both for WF, coinciding with HL, and BDS methods. As expected, the former method must be preferred to the latter one since it guarantees higher values of MI. Moreover, for both the receivers $I(A; B)$ surpasses the Holevo bound for small values of losses, while the situation becomes more critical for larger values. This can be better appreciated by the evaluation of the normalized information between Alice and Bob in the presence of collective attacks, that is
\begin{equation}\label{Kholevo}
K = \frac{I_{\rm p}(A; B)- \chi_{\rm p}(B; E)}{I_{\rm p}(A; B)} ,
\end{equation}
with p =WF or BDS. The experimental values of $|K|$ are shown in panels (b) and (c) of the same figure for WF and BDS receivers, respectively. In particular, full circles correspond to positive values of $K$ and open symbols to negative ones. We can easily observe that, at increasing losses, the values of $K$ are more negative. As already discussed about Fig.~\ref{individual}, we ascribe this behavior to the possible noise sources affecting the detection chain. In principle, further improvements of the setup and a more optimal choice of the parameters could lead to more stable results. 
\section{Conclusion}
In this work, we investigated the performance of SiPM detectors in the context of quantum communication. In particular, we considered a receiver exploiting the interference of a coherent signal with a LO before photodetection. As a figure of merit, we evaluated the MI in a BPSK protocol in the presence of losses and low intensity LO. Our results concerning the information transmission over the channel highlight its robusness with respect to losses taking also into account the non-unit visibility of the receiver.
This represents the basic element to develop more complex setups, such as the HYNORE receiver \cite{notarnicola}, to perform quasi-optimal discrimination strategies. In this view we plan to extend our analysis to higher order modulation formats, such as the amplitude PSK, and more advanced detection schemes.\\
Eventually, we studied a possible development towards CVQKD in the simplest case of a binary alphabet, where the obtained, though limited, promising results related to the evaluation of the KGR suggest further investigations with larger alphabets up to a continuous modulation.\\ 
From the practical point of view, to make the scheme more versatile, we plan to explore its extension in the telecom wavelengths range, for instance, realizing the optical part of the setup into optical fibers, while leaving the detection in the visible region and by using a sum-frequency generation process before the detection stage.


\section{Funding}
S. Cassina and A. Allevi acknowledge the support by PNRR D.D.M.M. 351/2022; M. Lamperti and A. Allevi acknowledge the support by PNRR D.D.M.M. 737/2021.

\section{Acknowledgments}
\begin{acknowledgments}
We thank Maristella Crotti (University of Insubria) for her assistance in the early stage of the experiment and Maria Bondani (Institute for Photonics and Nanotechnologies, CNR) for fruitful discussions.    
\end{acknowledgments}

\section{Disclosures}
The authors declare no conflicts of interest.

\section{Data Availability Statement}
The datasets generated and analyzed during the current study are available from the corresponding authors on reasonable request.

\clearpage


\begin{thebibliography}{99}
\bibitem{cariolaro} G. Cariolaro, \emph{Quantum Communications} (Springer Publishing Company, Incorporated, 2015).
\bibitem{flamini} F. Flamini, N. Spagnolo, and F. Sciarrino, 
``Photonic quantum information processing: a review,'' Rep. Prog. Phys. \textbf{82}, 016001 (2019).
\bibitem{grosshans} F. Grosshans and P. Grangier, ``Continuous Variable Quantum Cryptography Using Coherent States,'' Phys. Rev. Lett. \textbf{88}, 057902 (2002).
\bibitem{diamanti} E. Diamanti, H.-K. Lo, B. Qi, and Z. Yuan, `Practical challenges in quantum key distribution,'' Npj Quantum Inf. \textbf{2}, 16025 (2016).
\bibitem{olivares} S. Olivares, S. Cialdi, F. Castelli and M. G. A. Paris, ``Homodyne detection as a near-optimum receiver for phase-shift-keyed binary communication in the presence of phase diffusion,'' Phys. Rev. A \textbf{87}, 050303(R) (2013).
\bibitem{lvovsky} A. I. Lvovsky and M. G. Raymer, ``Continuous-variable optical quantum-state tomography,'' Rev. Mod. Phys. \textbf{81}, 299-332 (2009).
\bibitem{donati} G. Donati, T. J. Bartley, X.-M. Jin, M.-D. Vidrighin, A. Datta, M. Barbieri, and I. A. Walmsley, ``Observing optical coherence across Fock layers with weak-field homodyne detectors,'' Nature Commun. \textbf{5}, 5584 (2014).
\bibitem{IJQI17} A. Allevi, M. Bina, S. Olivares, and M. Bondani, ``Homodyne-like detection scheme based on photon-number-resolving detectors,'' Int. J. Quantum Inform. \textbf{15}, 1740016 (2017).
\bibitem{NJP19} S. Olivares, A. Allevi, G. Caiazzo, M. G. A. Paris, and M. Bondani, ``Quantum tomography of light states by photon-number-resolving detectors,'' New J. Phys. \textbf{21}, 103045 (2019).
\bibitem{OE17} M. Bina, A. Allevi, M. Bondani, and S. Olivares, ``Homodyne-like detection for coherent state-discrimination in the presence of phase noise
,'' Opt. Express, \textbf{25}, 10685-10692 (2017).
\bibitem{martinez} A. Martinez, “Spectral efficiency of optical direct detection,” J. Opt.
Soc. Am. B {\bf 24}, 739–749 (2007).
\bibitem{ribeiro} M. Cheraghchi and J. Ribeiro, “Improved upper bounds and structural
results on the capacity of the discrete-time Poisson channel,” IEEE
Trans. Inf. Theor. {\bf 65}, 4052–4068 (2019).
\bibitem{lukanowski} K. Łukanowski and M. Jarzyna, “Capacity of a Lossy Photon Channel
With Direct Detection,” IEEE Trans. Commun. {\bf 69}, 5059-5068 (2021).
\bibitem{jasc} M. N. Notarnicola and S. Olivares, ``Employing weak-field homodyne detection for quantum communications," arXiv:2405.14310 [quant-ph], 2024.
\bibitem{cattaneo} M. Cattaneo, M. G. A. Paris, and S. Olivares, ``Hybrid quantum key distribution using coherent states and photon-number-resolving detectors,'' Phys. Rev. A \textbf{98}, 012333 (2018).
 \bibitem{notarnicola23} M. N. Notarnicola, M. Jarzyna, S. Olivares, and K. Banaszek, ``Optimizing state-discrimination receivers for continuous-variable quantum key distribution over a wiretap channel,'' New J. Phys. \textbf{25}, 103014 (2023).
\bibitem{kennedy} R. S. Kennedy, ``A Near-Optimum Receiver for the Binary Coherent State Quantum
Channel,'' Quarterly Progress Report \textbf{108}, 219-225 (1973).
\bibitem{dimario} M. T. DiMario, and F. E. Becerra, ``Robust Measurement for the Discrimination of Binary Coherent States,'' Phys. Rev. Lett. \textbf{121}, 023603 (2018).
\bibitem{dimario1} M. T. DiMario, L. Kunz, K. Banaszek, and F. E. Becerra, ``Optimized communication strategies with binary coherent states over phase noise channels,'' npj Quantum Inf. \textbf{5}, 65 (2019).
\bibitem{OL19} G. Chesi, L. Malinverno, A. Allevi, R. Santoro, M. Caccia, and M. Bondani, ``Measuring nonclassicality with silicon photomultipliers,'' Opt. Lett. \textbf{44} 1371-1374 (2019).
\bibitem{JMO09} M. Bondani, A. Allevi, A. Agliati, and A. Andreoni, ``Self-consistent characterization of light statistics,'' J. Mod. Opt. \textbf{56}, 226-231 (2009).
\bibitem{cassina21} S. Cassina, A. Allevi, V. Mascagna, M. Prest, E. Vallazza, and M. Bondani, ``Exploiting the wide dynamic range of silicon photomultipliers for quantum optics applications,'' EPJ Quantum Technol. \textbf{8}, 4 (2021).
\bibitem{Pan} Z. Pan, K. P. Seshadreesan, W. Clark, M. R. Adcock, I. B. Djordjevic, J. H. Shapiro, and S. Guha, ``Secret-Key Distillation across a Quantum Wiretap Channel under Restricted Eavesdropping," Phys. Rev. Applied {\bf 14}, 024044 (2020).
\bibitem{PLAhomo} S. Olivares, A. Allevi, and M. Bondani, On the role of the local oscillator intensity in optical homodyne-like tomography,'' Phys. Lett. A \textbf{384}, 126354 (2020).
\bibitem{notarnicola} M. N. Notarnicola, M. G. A. Paris, and S. Olivares, ``Hybrid near-optimum binary receiver with realistic photon-number-resolving detectors,'' J. Opt. Soc. Am. B \textbf{40}, 705-714 (2023).
\bibitem{helstrom} C. W. Helstrom, \emph{Quantum Detection and Estimation Theory} (Elsevier, Academic Press, 1976).
\bibitem{bergou} J. A. Bergou, ``Discrimination of quantum states,'' J. Mod. Opt. \textbf{57}, 160-180 (2010).
\bibitem{izumi} S. Izumi, M. Takeoka, M. Fujiwara, N. Dalla Pozza, A. Assalini, K. Ema, and M. Sasaki, ``Displacement receiver for
phase-shift-keyed coherent states,'' Phys. Rev. A \textbf{86}, 042328 (2012).
\bibitem{muller} C. R. M$\ddot{\rm u}$ller and Ch. Marquardt, ``A robust quantum receiver for phase shift keyed signals,'' New J. Phys. \textbf{17}, 032003
(2015).
\bibitem{Olivares2021} S. Olivares, ``Introduction to generation, manipulation and characterization of optical quantum states," Phys. Lett. A {\bf 418}, 127720 (2021).
\bibitem{becerra} F. E. Becerra, J. Fan, G. Baumgartner, J. Goldhar, J. T. Kosloski, and A. Migdall, ``Experimental demonstration
of a receiver beating the standard quantum limit for multiple nonorthogonal state discrimination'' Nat. Photon. \textbf{7}, 147–152 (2013).
\bibitem{becerra1} F. E. Becerra, J. Fan, and A. Migdall, ``Implementation of generalized quantum measurements for unambiguous discrimination of multiple non-orthogonal coherent states,'' Nat. Commun. \textbf{4}, 2028 (2013).
\bibitem{cover} T. M. Cover, {\em Elements of information theory} (John Wiley \& Sons, 1999).
\bibitem{hama} $\rm https://www.hamamatsu.com/content/dam/hamamatsu-photonics/sites/documents/99\_SALES\_LIBRARY/ssd\\/s13360\_series\_kapd1052e.pdf$.
\bibitem{pirandola1} S. Pirandola et al., ``Advances in Quantum Cryptography,'' Adv. Opt. Photon. \textbf{12}, 1012–1236 (2020).
\bibitem{pirandola} S. Pirandola, ``Composable security for continuous variable quantum key distribution: Trust levels and practical key rates in wired and wireless networks," Phys. Rev. Research {\bf 3}, 043014 (2021).

\end{thebibliography}
\end{document}